\documentclass[11pt,a4paper]{article} 
\usepackage{jheppub}

\usepackage{amsfonts,,mathrsfs,latexsym,url,wasysym,float,slashed,xfrac,tipa}
\usepackage{mathtools}

\setcitestyle{square}

\newcommand{\be}{\begin{equation}}
\newcommand{\ee}{\end{equation}}

\newcommand{\ba}{\begin{eqnarray}}
\newcommand{\ea}{\end{eqnarray}}

\title{Emergent de Sitter Cosmology Near Black Hole Horizon}

\author{Ida M. Rasulian}

\affiliation{Department of Physics, Sharif University of Technology,\\Azadi Ave., Tehran, 1458889694, Iran}

\abstract{We propose an effective model for an exponentially expanding universe in the brane-world scenario. The setup consists of a 5D black hole and a brane close to the black hole horizon. In case the brane acquires a specific configuration, which we deduce from stability arguments, the induced metric outside the black hole horizon on the brane becomes de Sitter in static coordinates. Studying the Einstein equations perturbatively we find the effective gravity on the brane at this level and derive the 4D gravitational constant. Considering a homogeneous and isotropic fluid in the corresponding FLRW coordinates we  find that the bulk fluid density inside the brane, which has the same equation of state as the fluid on the brane, contributes to the energy density in the Friedmann equation and therefore in late time may be attributed to dark matter. Studying the stability of the setup we observe that the brane becomes stabilized, in the presence of matter on the brane, with a de Sitter length that is qualitatively of the order of Schwarzschild radius of the universe due to matter. We briefly discuss effects that can bound the de Sitter lifetime. In particular this model can provide a lifetime compatible with Trans-Planckian Censorship conjecture for the current de Sitter phase.}


\arxivnumber{2202.11596}

\begin{document}
\maketitle

\section{Introduction}
\label{intro}
It is by now common knowledge that the universe is currently in a state of accelerated expansion\cite{Riess}\cite{Perlmutter}. There is also substantial logical reasoning that the early universe has experienced an initial state of accelerated expansion\cite{Guth}\cite{Linde}. This motivates attempts for constructing de Sitter space-time in any possible framework. There has been a prolonged debate whether such a construction can be realized within the framework of string theory, (see for example \cite{Danielsson} for a review.) which lead to no-go theorems and conjectures that constrain such a construction\cite{Vafa1}\cite{Krishnan}\cite{Vafa2}\cite{Andriot} on the one hand and observational implications resulting from such constraints (for instance \cite{Kinney}\cite{Agrawal}) on the other.

Aside from issues regarding the construction of de Sitter space-time, there are also ongoing debates about its properties. A very important aspect of this is the lifetime of de Sitter space-time (see for instance \cite{Dvali}\cite{Bedroya1}\cite{Bedroya2}\cite{Freivogel}). Another important issue is the smallness of the current value for cosmological constant and the hierarchy between the value of Hubble parameter during inflation and late-time acceleration.

Due to these facts we tried to find a de Sitter construction in other frameworks, in this case the brane-world scenario. The idea behind the setup presented here originated from the similarities between black hole and de Sitter horizons, in the sense that the de Sitter horizon seems to be opposite to black hole horizon. Intuitively we interpret the de Sitter horizon as arising from a black hole living beyond it. In this framework the temperature of the constructed de Sitter space is in fact the near horizon temperature of the black hole on the observable part of the brane and the finite lifetime of de Sitter space is a result of collapse of matter towards the black hole beyond the horizon.
 
Here the resulting de Sitter length for the constructed de Sitter space-time is the geometric mean of two very small and very large length scales and it is conceivable that a small change in the infinitesimal length scale can lead to a large hierarchy in de Sitter length.

We should note \cite{Afshordi} where a setup with 5D black hole and DGP gravity was proposed for early universe. Other relevant work include \cite{Banerjee:2018qey}\cite{Banerjee:2019fzz}\cite{Dasgupta:2019gcd}.

In Sec. \ref{model} we introduce the setup along with a stability argument. In Sec. \ref{linear} we assume the brane is stabilized with some attributed de Sitter length and consider the presence of a stress tensor perturbation on the brane. Working to linear order in stress tensor we derive the 4D Friedmann equation on the brane and also derive the change in de Sitter length due to the presence of stress tensor. In Sec. \ref{lifetime} we discuss the de Sitter lifetime in this framework, and we continue to Sec. \ref{metricf} to discuss the issue of homogeneity and isotropy in this setup by studying the metric from the point of view of an in-falling observer on the brane. We conclude in Sec. \ref{conclude}.

\section{The Model}
\label{model}
\begin{figure}
\begin{center}
\includegraphics[scale=.26]{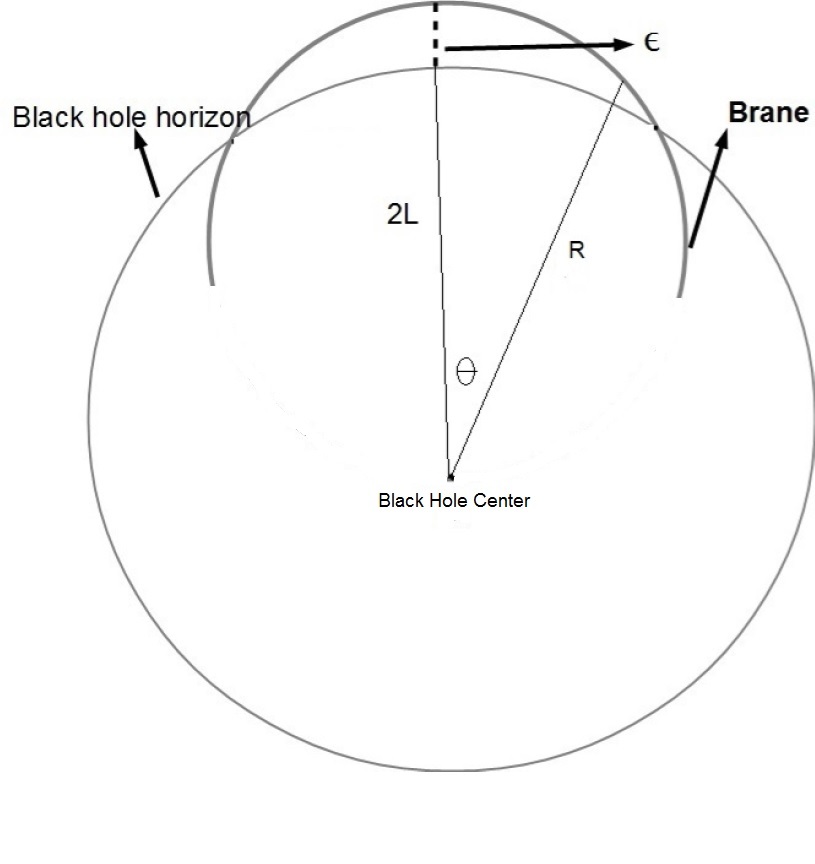}
\end{center}
\caption{The Setup Illustration}
\end{figure}
We propose an effective model for de Sitter space-time where 4D de Sitter universe is an emergent effect in a higher dimensional framework. In this model de Sitter space-time is not a vacuum solution of a string construction but an induced emergent effect from dynamics in higher dimensions. 

We consider a 5D Schwarzschild black hole and a brane close to its horizon. In case the brane acquires a specific configuration as in Fig. 1 the induced metric outside the black hole horizon on the brane becomes static de Sitter.

Assuming there is a black hole of mass $M$ and Schwarzschild radius $l_s^2\sim \kappa_5M$ in five dimensional space-time, where $\kappa_5$ is the 5D gravitational constant, the Schwarzschild metric is
\be\label{metric}
{\rm d}s^2=-\Big(1-\frac{l_s^2}{R^2}\Big){\rm d} t^2+\frac{{\rm d}R^2}{\Big(1-\frac{l_s^2}{R^2}\Big)}+R^2 {\rm d}\theta^2 +R^2 \sin ^2 \theta {\rm d}\Omega_2^2.
\ee
Moreover, there is a 3-brane such that the position of the brane is parameterized as
\be\label{position} R=2L\cos\theta.\ee
\be\label{defep} l_s=2L-\epsilon,\ee
where $\epsilon\ll L$.

The induced metric on the 3-brane outside the horizon, to the leading order in $\epsilon/L$, is given by
\be\label{induced-metric}
{\rm d}s_{ind}^2=-\Big(1-\frac{r^2}{l^2}\Big)\frac{\epsilon}{L}{{\rm d}t}^2+\frac{{\rm d}r^2}{\Big(1-\frac{r^2}{l^2}\Big)}+r^2 {\rm d}\Omega_2^2,
\ee
which is the 4D de Sitter metric in the static patch.
In the above, the de Sitter radial coordinate is $r=2L \sin\theta \approx 2L \theta$ and 
\be l\equiv 2\sqrt{\epsilon L}\approx \sqrt{2l_s\epsilon},\ee is the de Sitter horizon radius. 

We provide an explanation of why we think this setup can be stable. We first consider the case with no stress tensor on the brane. We show that the brane is stabilized such that $\epsilon=0$. As we discuss in section III we expect that in case we have a stress tensor on the brane it will be stabilized with some nonzero $\epsilon$ which is related to the amount of matter on the brane.
 
We first describe a model with a brane of positive tension. It is discussed numerically in \cite{Flachi} that if we have a black hole in the brane-world scenario on a flat brane and for some reason (perturbations) the black hole acquires a velocity perpendicular to the brane it modifies the shape of the brane and dissects a piece of it away from the so called mother brane. The result will be a black hole living on a closed brane.

The brane action can be written as 
\be
 S=-\sigma \int d^4 x \sqrt{-g},
\ee
where $\sigma$ is the brane tension and $g$ is the determinant of the induced metric. Parameterizing the brane's shape via the azimuthal angle $\theta$ the induced metric can be written as 
\be
{\rm d}s^2=-f(R)dt^2+f^{-1}(R)(R_{,\theta}d\theta+R_{,t}dt)^2+R^2 (d\theta^2+\sin^2\theta d\Omega_2^2) 
\ee
where $f(R)=1-\frac{l_s^2}{R^2}$ and the resulting Lagrangian will be 
\be
L=-\sigma(R \sin\theta)^2\sqrt{\Big | f(R) R^2-\frac{R^2 R_{,t}^2}{f(R)}+R_{,\theta}^2\Big |}.
\ee
We can see that in case of a static configuration and positive tension the solution $R=l_s \cos \theta$ provides a global maximum for the action. We can study the global maximum by solving for $f(R) R^2-\frac{R^2 R_{,t}^2}{f(R)}+R_{,\theta}^2=0$ assuming a static state and the result will be $R=l_s \cos \theta$ corresponding to $\epsilon=0$ in our model. The brane is trapped inside the black hole horizon. This provides an explanation of why a very small $\epsilon$ close to zero can be achieved in our setup. We generalize this setup to higher dimensions in Appendix B where we construct higher dimensional de Sitter space-times.

\section{Linearized 4D Gravity}
\label{linear}
In this section we study the effect of stress tensor at the perturbative level. We consider a configuration with de Sitter parameter $l_0$ and study effective gravity to linear order.

\subsection{Metric Near the Brane}
We first simplify the metric near the brane parameterized by de Sitter length $l_0$. We find it convenient to parameterize the normal coordinate on the brane in terms of the de Sitter length attributed to the observable part of the brane ($\theta\ll1$). The de Sitter length for the observable part of the brane is parameterized as $l=\sqrt{2l_s \epsilon}$ where $\epsilon$ is defined in \eqref{position} and \eqref{defep}. We foliate the normal direction to the brane by 3-branes with varying values of $\epsilon$ and therefore $l$. Near the brane, we parameterize the radial coordinate as
\be
R=(l_s+\epsilon)\cos \theta.
\ee
In this parameterization the metric \eqref{metric} is non-diagonal in $\theta$ and $\epsilon$.
It can be made diagonal by redefining the angular coordinate $\theta$. To find the new coordinate $\theta'$, we define $\theta=\theta(\theta',\epsilon)$. We have
\be 
d\theta=\partial_{\theta'}\theta d\theta'+\partial_\epsilon \theta d\epsilon
\ee
and 
\be 
dR=-(l_s+\epsilon)\sin \theta \partial_{\theta'}\theta d\theta'+\big(\cos \theta - (l_s+\epsilon)\sin \theta \partial_\epsilon \theta\big)d\epsilon
\ee
Then, the non-diagonal part of the metric is
\be
{\rm d}s^2 \supset 2d\theta' d\epsilon \partial_{\theta'}\theta \Big[(l_s+\epsilon)^2\cos^2\theta \partial_\epsilon \theta +\frac{(l_s+\epsilon)^2\sin^2\theta \partial_\epsilon \theta-(l_s+\epsilon)\sin\theta \cos\theta}{1-\frac{l_s^2}{R^2}}\Big].
\ee
The off-diagonal component is vanishing for
\be 
\frac{2\partial_\epsilon \theta}{\sin 2\theta}=\frac{l_s+\epsilon}{2l_s \epsilon+\epsilon^2},
\ee
from which, after $\epsilon$-integration, we find
\be
\tan \theta=c(\theta')\sqrt{(l_s+\epsilon)^2-l_s^2}, 
\ee
where $c(\theta')$ is an arbitrary function of $\theta'$.
By demanding that $\theta(\theta',\epsilon_0)=\theta'$ (where $\epsilon_0$ is related to $l_0$), we find
\be 
\tan \theta =\frac{\sqrt{(l_s+\epsilon)^2-l_s^2}}{l_0} \tan \theta'.
\ee
For very small $\theta$ and $\theta'$ that is relevant for the observable part of the brane this is equivalent to
\be 
\theta \approx \frac{l}{l_0} \theta'.
\ee
We define 
\be r = l_s\sin\theta'.\ee
Therefore for instance we have 
\be 
-g_{tt}=1-\frac{l_s^2}{(l_s+\epsilon)^2\cos^2 \theta} \approx \frac{-l_s^2 \theta^2 + l^2}{l_s^2}=\frac{l^2}{l_0^2}\frac{2\epsilon_0}{l_s}\big(1-\frac{r^2}{l_0^2}\big).
\ee
Consequently, the 5D metric close to the 3-brane with de Sitter length $l_0$, in the limit of small $\theta$ and $\epsilon/l_s$, is
\ba\label{background-metric}
{\rm d}s^2&=&\frac{l^2}{l_0^2}\Big(-f(r)dt'^2+\frac{dr^2}{f(r)}+r^2 d\Omega_2^2\Big)+dl^2,
%&=& \frac{l^2}{l_0^2}g_{\mu\nu}dx^\mu dx^\nu+dl^2, 
\ea
where
\be
f(r)=1-\frac{r^2}{l_0^2}. 
\ee
and 
\be 
dt'=\sqrt{\frac{2\epsilon_0}{l_s}}dt
\ee
in the scaled time on the brane. In passing we emphasize that here $l$ is the normal coordinate to the brane. 

\subsection{Linearized Einstein Equations}
We consider the following form of perturbations around the background metric \eqref{background-metric}\cite{Csaki}
\be\label{perturb}
{\rm d}s^2=\frac{l^2}{l_0^2}(g_{\mu\nu}+h_{\mu\nu})dx^{\mu}dx^{\nu}+dl^2,
\ee
where we choose a gauge such that $h_\mu^\mu=0$. For the rest of analysis, we take the conformal FLRW form of the induced metric
\be 
{\rm d}s_{in}^2=g_{\mu\nu}dx^{\mu}dx^{\nu}=\frac{l^2}{\tau^2}(-d\tau^2 + d\vec{\rm x}^2),
\ee
where $\tau$ is conformal time.

In the presence of a stress tensor, the position of the brane changes\footnote{As explained previously we quantify the position of the brane by the parameter $l$ which is related to the de Sitter length of the observable part of the brane.} and the displacement is given by the brane bending parameter $\zeta(x)$ as \cite{Csaki,Seahra}
\be
l=l_0+\zeta(x^\mu).
\ee
We can choose a coordinate system, $\bar{x}^{\mu}$ and $\bar{l}$, such that the brane is located at $\bar{l}=l_0$ and the metric takes the same form as \eqref{perturb}. The new coordinates are
\ba 
\bar{l}&=&l-\zeta(x^\mu),
\\ \bar{x}^\mu &=& x^\mu-\zeta^\mu+\partial^\mu \zeta \int^l dl' \frac{\bar{\tau}^2}{l^2}.
\ea
Then, the metric perturbations in the old and new coordinates are related as
\be\label{rel}
h_{\mu\nu}=\bar{h}_{\mu\nu}-D_\mu \zeta_\nu-D_\nu \zeta_\mu-2D_\mu \partial_\nu \zeta \frac{\bar{\tau}^2}{l} -2\eta_{\mu\nu}\frac{\zeta}{l}.
\ee
For future reference we perform a coordinate redefinition to transform the metric to a conformally $Z_2$-symmetric form
\be 
ds^2=e^{\frac{2z}{l_0}}\big(\frac{l_0^2}{\tau^2}(\eta_{\mu\nu}+h_{\mu\nu})dx^\mu dx^\nu +dz^2\big),
\ee
where we have $\frac{l}{l_0}=e^{\frac{z}{l_0}}$ and we assume the $Z_2$-breaking effect in this geometry only lies in the conformal factor.

In this coordinate \eqref{rel} can be written as
\be\label{relz}
h_{\mu\nu}=\bar{h}_{\mu\nu}-D_\mu \zeta_\nu-D_\nu \zeta_\mu-2D_\mu \partial_\nu \zeta \frac{\bar{\tau}^2}{l_0}e^{-\frac{\bar{z}}{l_0}} -2\eta_{\mu\nu}\frac{\zeta}{l_0}e^{-\frac{\bar{z}}{l_0}}.
\ee

The Einstein equations can be written as\cite{Randall}
\be\label{gn}
2G_{55}=-R^{(4)}+\frac{3}{l_0}\partial_z h+\frac{12}{l_0^2}=2\kappa_5 T_{zz},
\ee
\be\label{gt}
G_{\mu\nu}=G^{(4)}_{\mu\nu}+\frac{3}{l_0^2}g_{\mu\nu}+\frac{3l_0}{2\tau^2}\partial_z(h \eta_{\mu\nu}-h_{\mu\nu})+\frac{l_0^2}{2\tau^2}\partial^2_z(h \eta_{\mu\nu}-h_{\mu\nu})=\kappa_5 T_{\mu\nu}
\ee
and 
\be 
2G_{5\mu}=\partial_z\big(\frac{4}{\tau}h_{0\mu}+\frac{1}{\tau}h \delta^0_\mu-\partial_\mu h+\frac{l_0^2}{\tau^2}\partial^\alpha h_{\alpha \mu}\big)=2\kappa_5 T^5_\mu.
\ee
Taking a trace over \eqref{gt} and using \eqref{gn} we find
\be 
\frac{3}{2l_0}\partial_z h+\frac{3}{2}\partial^2_z h=\kappa_5 (T-2T_{zz}),
\ee
where we have not included the factor of $e^{\frac{2z}{l_0}}$ in $T$. On the other hand from energy conservation and assuming for simplicity $T_{\mu 5}=0$ we find 
\be\label{conszz}
\partial_z T_{zz}+\frac{2}{l_0} T_{zz}=\frac{1}{l_0} T.
\ee
Therefore we have 
\be 
\frac{3}{2l_0}\partial_z(h+l_0 \partial_z h)=\kappa_5 l_0 \partial_z T_{zz}
\ee
and 
\be 
\frac{3}{2l_0}(h+l_0 \partial_z h)=\kappa_5 l_0 T_{zz}.
\ee
In the $\bar{h}$ frame we have $\langle \partial_z \bar{h}\rangle =0$\footnote{In our notation $[A]$ represents the jump in $A$ and $\langle A\rangle$ represents its mean value.} and 
\be 
\bar{h}=\frac{2\kappa_5 l_0^2}{3}\langle T_{zz}\rangle.
\ee
In $h$ frame we have $h=0$ and 
\be 
\langle \partial_z h\rangle =\frac{2\kappa_5 l_0}{3}\langle T_{zz} \rangle.
\ee
From these considering \eqref{relz} we find an equation for $\zeta$
\be\label{zetaeq}
\square^{(4)}\zeta+\frac{4\zeta}{l_0^2}=\frac{2}{3}\kappa_5 l_0 \langle T_{zz} \rangle.
\ee
Assuming $\langle \partial_z T_{zz} \rangle=0$ we find from \eqref{conszz} 
\be\label{tzz} 
\langle T_{zz} \rangle=\frac{\langle T \rangle}{2}.
\ee
Next we return to Einstein equations in $\bar{h}$ frame. Considering the delta function type contribution of brane stress tensor we find from \eqref{gt} 
\be 
\frac{l_0^2}{2\tau^2}\partial^2_z(\bar{h}\eta_{\mu\nu}-\bar{h}_{\mu\nu})=\kappa_5 S_{\mu\nu} \delta(z)+finite.
\ee
Therefore we have\cite{Israel}
\be 
\frac{l_0^2}{2\tau^2}[\partial_z(\bar{h}\eta_{\mu\nu}-\bar{h}_{\mu\nu})]=\kappa_5 S_{\mu\nu}
\ee
Considering the jump in \eqref{gt} we find
\be\label{jumptrans}
[T_{\mu\nu}]=\frac{3S_{\mu\nu}}{l_0}.
\ee
We further need the finite contribution to $\partial^2_z$ part of \eqref{gt} to complete the calculation. Taking a derivative with respect to $z$ in \eqref{gt} and considering the mean value we find 
\be 
\frac{3l_0}{2\tau^2} \langle\partial^2_z(\bar{h} \eta_{\mu\nu}-\bar{h}_{\mu\nu})\rangle=\kappa_5 \langle \partial_z T_{\mu\nu}\rangle,
\ee
which is zero in case $\langle \partial_z T_{\mu\nu}\rangle=0$ that we infer from approximate $Z_2$-symmetry.

We finaly find from the mean value of \eqref{gt} 
\be\label{4deinstein}
G^{(4)}_{\mu\nu}+\frac{3}{l_0^2}g_{\mu\nu}=\kappa_5 (T^{-}_{\mu\nu}+\frac{3S_{\mu\nu}}{2l_0}),
\ee
where we used \eqref{jumptrans}. Here we have separated the effect of $T^{-}_{\mu\nu}$ which is the bulk stress tensor below the brane. From here we find the effective 4D gravitational constant as 
\be\label{4dg}
\kappa_4=\frac{3\kappa_5}{2l_0}.
\ee
Given that currently $l_0^{-1}\sim10^{-33}$eV we conclude that the 5D quantum gravity scale is as low as
$m_5\sim100$ MeV. We provide a justification of why the effective 4D cut-off on the brane can be much higher. The scale of energy density in the bulk for which 5D quantum gravity becomes important is $m_5^5$. This corresponds to energy scales of order $l_0 m_5^5$ on the brane. Considering this to be of order $m_4^4$ where $m_4$ is the corresponding cut-off on the brane we find $m_4 \sim 10^6$ TeV.

Confining to the case of a homogeneous and isotropic fluid it would be interesting to study the effect on the position of the brane parameterized by $\zeta(x^\mu)$. In this case we find from \eqref{zetaeq} that
\be 
-\ddot{\zeta}-\frac{3}{l_0}\dot{\zeta}+\frac{4}{l_0^2}\zeta=\frac{2}{3}\kappa_5 l_0 \langle T_{zz}\rangle=\frac{1}{3}\kappa_5 l_0 \langle T \rangle=\frac{1}{3}\kappa_5 l_0 (T^-+\frac{3S}{2l_0}),
\ee
where we again used \eqref{jumptrans} from which we also expect the same equation of state for the fluid corresponding to $T_{\mu\nu}$ and $S_{\mu\nu}$. Considering this to be $p=\omega \rho$ we find for generic $\omega$
\be 
\zeta \propto e^{-\frac{3(1+\omega)}{l_0}}
\ee
and 
\be 
\frac{(1-3\omega)(4+3\omega)}{l_0^2}\zeta=\frac{-1+3\omega}{3}\kappa_5 l_0 (\rho^-+\frac{3}{2l_0}\rho_S),
\ee
or
\be 
(4+3\omega)\zeta = -\frac{1}{3}\kappa_5 l_0^3(\rho^-+\frac{3}{2l_0}\rho_S)=-\frac{\kappa_5 l_0^2}{2}\rho_{tot}.
\ee
Here we defined $\rho_{tot}=\rho_S+\frac{2l_0}{3}\rho^-$, where $\rho_S$ is the energy density on the brane and $\rho^-$ is the energy density below the brane. We recognize $\rho_{tot}$ using \eqref{4deinstein} to be the effective matter density on the brane that has a contribution from the bulk as well. We may wish to attribute this extra invisible source of matter to dark matter in the matter-dominated phase, although this needs further study.

We provide a qualitative description for the consequences of this equality for brane stability. In case there is no stress tensor on the brane and in the bulk, we expect it to be stabilized with $l\approx 0$. When we turn on the stress tensor the position changes and the brane will be stabilized with a new $l$ and infinitesimally we can write 
\be 
dl \propto - \kappa_5 l^2 d\rho_{tot},
\ee
or upon integration
\be 
\frac{1}{l}\propto \kappa_5 \rho_{tot}.
\ee
From this using \eqref{4dg} we find that the contribution of dark energy part, in an approximately de Sitter phase, will be proportional to the contribution from effective matter i.e.
\be 
\frac{3}{\kappa_4 l^2} \propto \rho_{tot}.
\ee
\section{de Sitter Lifetime}
\label{lifetime}
An intriguing fact about black holes is that Hawking radiation can be approximated by that of a black body. An observer living close to the horizon sees a hot bath of particles with a temperature that is specified by the surface gravity.

Using the 5D black hole metric \eqref{metric} if we define $R=l_s+\frac{\rho^2}{2l_s}$ and $\tau=\frac{t}{l_s}$ we have
\be
ds^2=-\rho^2 d\tau^2+d\rho^2+....
\ee
From here substituting $\rho=\sqrt{l_s(R-l_s)}$ and $R=2L \cos \theta$ on the brane, we can read the temperature using $T_b=\frac{1}{2\pi \rho}$
\be
T_b= \frac{\hbar c}{2\pi k_B l\sqrt{1-(\frac{r}{l})^2}},
\ee
with $l=2\sqrt{L\epsilon}$.
If we assume we live close to the center $r\ll l$ the temperature would be
\be
T_b=\frac{\hbar c}{2\pi k_B l}.
\ee
This can be interpreted as de Sitter temperature.

Specifically Hawking radiation in the background of 5D black hole results in a temperature equal to $\frac{1}{2\pi l}$ close to the brane which corresponds to energy density of order $\rho \propto \frac{1}{l^5}$ from the point of view of near horizon observer in the bulk. According to \eqref{jumptrans} this results a density of order $\rho_S \propto \frac{1}{l^4}$ on the brane, which corresponds to Hawking radiation with temperature $\frac{1}{2\pi l}$.

In this section we first discuss the effect of black hole evaporation. We then consider the effect of black hole collapse on de Sitter lifetime.

Studying the black hole evaporation from 5D point of view, in case we assume matter fields on the brane interact with matter fields in the bulk only gravitationally, the evaporated energy from the black hole would not effect the observable brane and thereby has negligible impact on de Sitter lifetime. 

From the point of view of matter fields on the brane the quantum stress tensor as studied in for instance \cite{Markkanen} has no flux and sheer and therefore semi-classically there is no flux of energy from the black hole to the observable part of brane. Therefore we conclude that semi-classical effect is negligible in de Sitter lifetime in this setup. We speculate that in case we have a very small de Sitter length, of order 5D Planck's length, the 5D stress tensor will be of order $\frac{1}{l_5^5}$ and 5D quantum gravity effects become important.

Next we argue that there are effects other than semi-classical evaporation that we should take into account. During some de Sitter phase with parameter say $\epsilon_1$ (defined as $l_s=2L-\epsilon_1$) the effect other than black hole evaporation which can happen in a faster rate is the in-fall of matter towards the black hole. We provide an approximation for this timescale by considering the time it takes for a photon to travel from a point $r_0$ to the vicinity of the horizon with proper distance $l_5$. Here $l_5$ is the 5D Planck's length defined as
\be
l_5=(\kappa_5 \hbar)^{\frac{1}{3}}=(l_0 (l^0_p)^2)^{\frac{1}{3}}.
\ee
Here we used \eqref{4dg} and $l_p^2=\kappa_4 \hbar$.\footnote{In this notation the $'0'$ index corresponds to the current value of quantities.} We first find the $\delta r$ corresponding to proper distance $l_5$ from the horizon
\be
\int_{l_1-\delta r}^{l_1} \frac{dr}{\sqrt{1-\frac{r^2}{l_1^2}}}=\sqrt{2l_1 \delta r} =l_5
\ee
and we find 
\be
\delta r=\frac{l_5^2}{2l_1}.
\ee
Then we find the approximate time scale for the in-fall of matter towards the black hole (Here we assume collapse takes place when matter has fallen to the vicinity of the horizon with proper distance $l_p$ as in \cite{Greenwood}, which is in contrast to the relevant paper \cite{Vachaspati}.)
\be
T\approx \int_{r_0}^{l_1-\delta r} \frac{dr}{1-\frac{r^2}{l_1^2}}\approx l_1 Log(\frac{l_1^3}{l_0 (l^0_p)^2}).
\ee
We mention that this calculation is valid only for $l_1>(l_0 (l^0_p)^2)^{1/3}$. Otherwise the observable part of the brane is inside the quantum membrane with proper distance of 5D Planck's length from the horizon and this classical approximation does not hold. In case the inflationary phase happens with a de Sitter parameter of order $l_5$ then the universe in this phase is inside the quantum membrane of 5D black hole.
\section{Metric from the Point of View of a Massive In-falling Observer} 
\label{metricf}
Here we want to study the metric from the point of view of a massive observer that moves on a geodesic originating from the center of static de Sitter space-time and calculate deviations from exponentially expanding FLRW metric. 

Considering radial motion for simplicity we define the radial position of the observer to be $r_0$. We can find from geodesic equation
\be\label{geo}
\frac{d^2 r_0}{d\tau^2}+\frac{r_0}{l^2}\Big(\frac{\dot{r}_0^2}{1-\frac{r_0^2}{l^2}}-\dot{t}^2(1-\frac{r_0^2}{l^2})\Big) =0,
\ee
where dot represents derivative with respect to the affine parameter. We have for the affine parameter $\tau$
\be\label{tau}
-d\tau^2=-(1-\frac{r_0^2}{l^2}) dt^2+\frac{dr^2}{(1-\frac{r_0^2}{l^2})},
\ee
therefore \eqref{geo} can be written as
\be
\frac{d^2 r_0}{d\tau^2}=\frac{r_0}{l^2},
\ee
or using only positive exponential term
\be\label{rdot}
\dot{r_0}=\frac{r_0}{l}.
\ee
Using \eqref{tau} again we find
\be
\dot{t}=\frac{1}{1-\frac{r_0^2}{l^2}}.
\ee
Dividing \eqref{rdot} by this we find
\be 
\frac{dr_0}{dt}=\frac{r_0}{l}\big(1-\frac{r_0^2}{l^2}\big)
\ee
and solving we find
\be
t=l Ln \frac{r_0}{l}-\frac{l}{2}Ln(1-\frac{r_0^2}{l^2}).
\ee
We later need this in terms of FLRW time which can be written as
\be
T=t+\frac{l}{2}Ln \big(1-\frac{r_0^2}{l^2}\big) =l Ln \frac{r_0}{l}.
\ee
We see from these equations that for $r_0$ to start from $r_0\approx 0$ and approach a point comparable to $l$ an infinite static and FLRW time is required. So for later use we assume $\frac{r_0}{l}$ remains very small.

We relate the new coordinates centered at $r_0$ to the old coordinates centered at the origin of static de Sitter patch. We have
\be
r^2=r_0^2+r'^2+2r_0 r' \cos \phi ',
\ee
where $r'$ is the new radial coordinate on the brane measured by the observer centered at $r=r_0$ and $\phi'$ is the corresponding angular coordinate. We also have  
\be 
dr^2+r^2 d\Omega_2^2=(dr')^2+(r')^2 d(\Omega'_2)^2.
\ee
From these and using the transition to FLRW coordinates in the usual way via
\be 
r'=R e^{\frac{T}{l}}
\ee
and 
\be
t=T-\frac{l}{2}Ln\big(1-\frac{r'^2}{l^2}\big),
\ee
we find a modified FLRW metric for the observer who is slightly removed from the center keeping only linear terms in $\frac{r_0}{l}$.
\begin{multline}
ds^2=-\Big(1-\frac{2R \, r_n}{l^2}\cos \phi ' e^{\frac{2T}{l}}\frac{1+2\frac{r'^2}{l^2}}{(1-\frac{r'^2}{l^2})^2}\Big)dT^2+e^{\frac{2T}{l}}\Big(1+\frac{2R \, r_n}{l^2}\cos \phi ' e^{\frac{2T}{l}}\frac{1+2\frac{r'^2}{l^2}}{(1-\frac{r'^2}{l^2})^2}\Big)dR^2\\+2dT dR \frac{r_n}{l} \cos \phi' \frac{\frac{r'^2}{l^2}}{(1-\frac{r'^2}{l^2})^2} e^{\frac{2T}{l}}\big(5+3\frac{r'^2}{l^2}\big) -2d\phi ' dR \, r_n \sin \phi' \frac{\frac{r'^2}{l^2}}{1-\frac{r'^2}{l^2}}e^{\frac{2T}{l}}\\-2d \phi' dT r_n \sin \phi' \frac{\frac{r'^3}{l^3}}{1-\frac{r'^2}{l^2}}e^{\frac{T}{l}}+R^2 e^{\frac{2T}{l}}d{\Omega'_2}^2. 
\end{multline}
Here we have chosen $T_{now}=0$ and $r_n$ is the present value of $r_0$. It is understood that $r'$ in this equation should be set to its FLRW value $R e^{\frac{T}{l}}$. It is evident from this result that a non-zero $r_n$ breaks isotropy by terms that depend on the angular coordinate $\phi'$.
\section{Conclusions}
\label{conclude}
An interesting observation regarding this work is relating the lifetime of the constructed de Sitter space-time to classical and quantum gravity effects. Our result for semi-classical de Sitter evaporation rate implies that semi-classical evaporation has negligible effect on de Sitter lifetime in this setup.

Another important aspect of this model is the appearance of a timescale similar to scrambling time in predicting the de Sitter lifetime. This happens because the escape of matter from de Sitter horizon (whilst entering the black hole horizon) increases the black hole mass and therefore radius which ultimately distorts the resulting de Sitter patch and provides an approximated de Sitter lifetime.

We found a bound on the lifetime of this effective de Sitter space-time that respects the Trans-Planckian censorship conjecture for large de Sitter length.
After this timescale the de Sitter space-time is distorted and our prediction is that this in-fall of matter and decrease of $\epsilon$ continues until $\epsilon$ becomes small. When we approach small de Sitter length quantum gravity effects will become important.

Regarding the stability of the model we observed qualitatively at the perturbative level that stability requires a proportionality between two different species of energy i.e. matter and dark energy density. Also the presence of an effect due to bulk fluid in the Friedmann equation may be related to dark matter. 

This setup provides a 4D emergent space-time which is approximately de Sitter, as the derived metric is in the static patch and depends on the observer. From this aspect we tried to provide a picture for the deviation from homogeneity and isotropy as seen by a massive free-falling observer on the brane, that starts its journey close to the center of the de Sitter patch ($\frac{r}{l}\ll 1$). We observe that this deviation can be negligible given a small enough initial $\frac{r}{l}$. 

A possible explanation for this small value can be achieved in case we have a complete cosmological description in this framework that also includes a period of early time inflation. During this time the 4D universe is expected to be characterized by a small de Sitter length $l_i$ and if the de Sitter radius changes later to say the current value $l_0$ the fine-tuning in the initial value of $\frac{r}{l}$ can be achieved for $\frac{l_i}{l_0}\ll 1$. This has a resemblance to the way standard inflation solves the homogeneity and isotropy problem.

There are many open issues regarding this construction. The most important one is understanding the induced gravity on the brane and stability non-perturbatively. At the next level the cosmic history that the model predicts needs to be studied in detail. In order to achieve this end one needs to also consider the effect of collapse and 5D quantum gravity that make the de Sitter length dynamical. A possible obstacle is that 5D quantum gravity becomes important at early time for small de Sitter length (of order the 5D Planck's length). 

This is due to the fact that for such a small de Sitter length the observable part of the brane is inside the quantum membrane (of proper Planck's distance from the horizon) near black hole horizon. For a free-falling observer we expect no new physics in this region, but since the brane is stabilized and not free-falling the Unruh temperature on the brane becomes larger than 5D Planck's scale and one might expect effects of quantum gravity to become important\footnote{This brings the question: Is quantum gravity observer-dependent?}. 

\section*{Acknowledgements}
This work has been done under the supervision of Dr. Mahdi Torabian and Dr. Hessamaddin Arfaei, towards partial fulfillment of the requirements of the degree of Doctor of Philosophy at Sharif University of Technology. The author is grateful to Dr. Mahdi Torabian for fruitful discussions and comments
%% The Appendices part is started with the command \appendix;
%% appendix sections are then done as normal sections
%% \appendix
\appendix
\section{Kerr and AdS-Schwarzschild Black Holes}
Here we study the induced metric for AdS-Schwarzschild and Kerr black holes. For the first case we have the background metric
\be
ds^2=-\Big(1-\frac{l_s^2}{R^2}+\frac{R^2}{a^2}\Big) dt^2+ \Big(1-\frac{l_s^2}{R^2}+\frac{R^2}{a^2}\Big)^{-1}dR^2 +R^2 d\Omega_3^2
\ee
and the induced metric assuming $L \ll a$ will be 
\be
ds_{in}^2=-\Big(\frac{4L^2}{a^2}+\frac{\epsilon}{L}\Big)\Big(1-\frac{r^2}{l^2+\frac{16L^4}{a^2}}\Big) dt^2+ \Big(1-\frac{r^2}{l^2+\frac{16L^4}{a^2}}\Big)^{-1}dr^2+r^2 d\Omega_2^2.
\ee
In case we have a 5D Kerr black hole with metric \cite{Myers}
\ba
ds^2= -dt^2+\psi dR^2+\rho^2 d\theta^2+\sin^2 \theta(R^2+a^2)d\phi^2+\Delta(dt+a \sin^2 \theta d\phi)^2\cr +R^2 \cos^2 \theta d\chi^2
\ea
and definitions
\be
\rho^2=R^2+a^2 cos^2 \theta 
\ee
\be
\Delta=\frac{l_s^2}{\rho^2} 
\ee
\be
\psi=\frac{\rho^2}{R^2+a^2-l_s^2},
\ee
we find the induced metric
\begin{multline}
ds_{in}^2=-\frac{1-\frac{r^2(1+\alpha^2 \sin^2 \theta)}{l^2}}{1+\alpha^2}dt^2+\frac{1+\alpha^2 sin^2  \theta+\alpha^2 \frac{r^2}{l^2}\cos^2 \theta}{1-\frac{r^2}{l^2}} dr^2 +r^2(1+\alpha^2 \sin^2\theta)d\theta^2\\+r^2 \sin^2 \theta d\phi^2- \frac{2\alpha r^2}{(1+\alpha^2)l}\sin^2 \theta d\phi dt+2r \sin \theta \cos \theta \alpha^2 dr d\theta,
\end{multline}

where we defined $\alpha=\frac{a}{L}$. In case $\alpha\ll 1$ this can be written as 
\begin{multline}
ds_{in}^2=-\big(1-\frac{r^2}{l^2}\big)dt^2+ \big(1-\frac{r^2}{l^2}\big)^{-1}dr^2 +r^2 d\theta^2+ r^2 \sin^2 \theta d\phi^2-\frac{2\alpha r^2}{l}\sin^2 \theta dt d\phi,
\end{multline}

which is the metric for a rotating de Sitter space-time.
%% \section{}
%% \label{}

%% If you have bibdatabase file and want bibtex to generate the
%% bibitems, please use
%%
%%  \bibliographystyle{elsarticle-harv} 
%%  \bibliography{<your bibdatabase>}

%% else use the following coding to input the bibitems directly in the
%% TeX file.
\section{Generalization to Higher Dimensions}
We consider a Schwarzschild black hole in $n+3$ dimensions
\be 
ds^2=-(1-\frac{l_s^n}{R^n})dt^2+\frac{dR^2}{1-\frac{l_s^n}{R^n}}+R^2 d\Omega_{n+1}^2
\ee
and consider a brane in its vicinity with position parameterized as $R=R(\theta)$. Then considering only the $\theta$-dependent part we have 
\be 
\sqrt{-g}=\sin^n \theta \sqrt{R^2(\theta)+R'^2(\theta)-\frac{l_s^n}{R^{n-2}(\theta)}}.
\ee
Requiring this to vanish we find for $R(\theta)$
\be 
R(\theta)=l_s \cos^{\frac{2}{n}}(\frac{n \theta}{2}).
\ee
Considering a small deformation parameterized with $\epsilon$ as
\be 
R(\theta)=(l_s+\epsilon)\cos^{\frac{2}{n}}(\frac{n \theta}{2}),
\ee
we derive the induced metric on the brane for small $\theta$. We find 
\be 
ds_{in}=-\frac{n^2 l^2}{4l_s^2}(1-\frac{r^2}{l^2})dt^2+\frac{dr^2}{1-\frac{r^2}{l^2}}+r^2 d\Omega_{n}^2,
\ee
where we have $r=l_s \theta$ and 
\be 
l^2=\frac{4}{n}l_s \epsilon.
\ee

\end{document}